\documentclass[12pt]{article}

\usepackage{graphicx}

\begin{document}

\vskip 0.5cm \centerline{\bf Exclusive diffraction and Pomeron trajectory}
\centerline{\bf in {\em ep} collisions}
\vskip 0.3cm \centerline{
S.~Fazio$^{a,b\diamond}$ 
and L.~Jenkovszky$^{c,d\star}$}

\vskip 1cm

\centerline{ $^a$ \sl Dipartimento di Fisica, Universit\'a della
Calabria} \centerline{ \sl Istituto Nazionale di Fisica Nucleare,
Gruppo collegato di Cosenza} \centerline{  \sl I-87036 Arcavacata
di Rende, Cosenza, Italy} 
\centerline{$^{b}$  \sl Joint Institute for Nuclear Research, 141980 - Dubna, Moscow Region, Russia}
\centerline{$^c$ \sl Bogolubov Institute
for Theoretical Physics, National Academy of Sciences of Ukraine}
\centerline{\sl Kiev-143, 03680 Ukraine} \centerline{$^{d}$  \sl
KFKI, RMKI, Budapest, Hungary}

\vskip 0.1cm

\begin{abstract}
The exclusive diffractive production of vector mesons and real photons in $ep$ collisions has been studied at HERA in a wide kinematic range. Here we present the most recent experimental results together with a Regge-type model. We deduce the Pomeranchuk trajectory (Pomeron) by analyzing the HERA data on deeply virtual Compton scattering (DVCS), and then discuss its basic properties, namely its apparent \textquotedblleft hardness\textquotedblright
and its \textquotedblleft non-flat\textquotedblright behavior, different from the claims of some authors. 
\end{abstract}

\vskip 0.1cm

$
\begin{array}{ll}
^{\diamond}\mbox{{\it e-mail address:}} &
   \mbox{fazio@fis.unical.it} \\
^{\star}\mbox{{\it e-mail address:}} &
   \mbox{jenk@bitp.kiev.ua} \\

\end{array}
$

\section{Introduction}
The diffractive scattering is a process where the colliding particles scatter at very small angles and without any color flux in the final state. This involves a propagator carrying the vacuum quantum numbers, called Pomeron and described, in the soft regime, within the Regge theory. Since the first operation period in 1992, ZEUS and H1, the two experiments dedicated to the DIS physics at HERA, observed that a big amount $(\sim 10 \%)$ of lepton-proton DIS events had a diffractive origin opening a new area of studies in diffractive production mechanism, providing a hard scale which can be
varied over a wide range and therefore it is an ideal testing for QCD models.

In particular, the diffractive production of Vector Mesons (VMs) and real photons in $ep$ collisions allows to study the transition from the soft to the hard regime in strong interactions. The hard regime (high energy and low Bjorken-$x$) is dominated by the exchange of a hard Pomeron sensitive to the gluon content and well described by perturbative QCD (pQCD), while in the soft regime (low-$x$) the interaction is well described within the Regge phenomenology. Indicating with $Q^2$ the virtuality of the exchanged photon and with $M^2$ the square mass of the produced VM, HERA data suggested a universal hard scale, $Q^2+M^2$, for the diffractive exclusive pruduction of VMs and real photons, which indicates the transition from the soft to the hard regime.
 

\section{$Q^2$ and $W$ dependence of the cross section}

\begin{figure}[htbp]
\centering
\includegraphics[width=0.7\textwidth,angle=0]{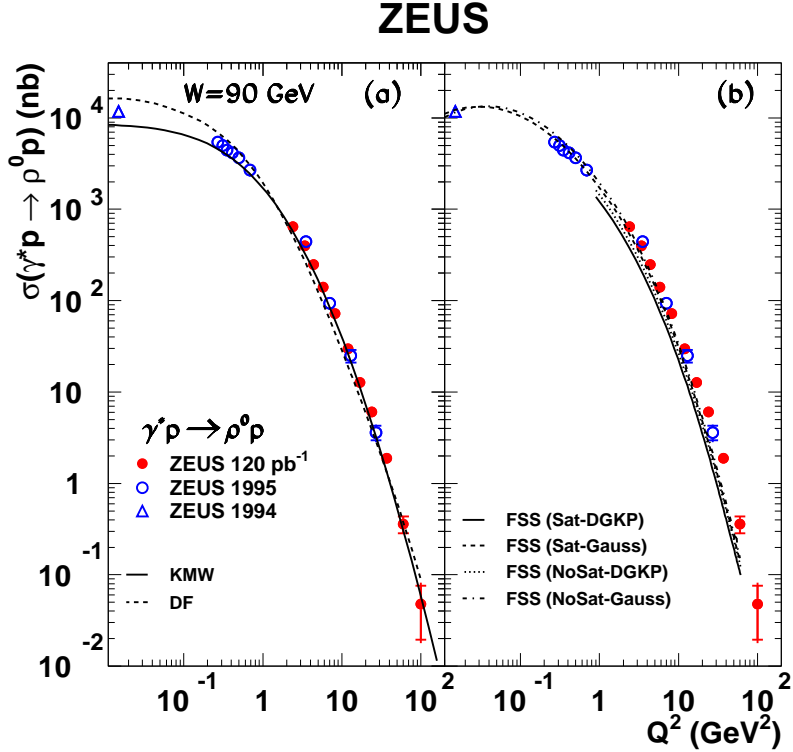}
\caption{The $\gamma^*p\rightarrow\rho^0p$ cross section as a function of $Q^2$ measured at $W=90\;GeV^2$ and comared in (a) and (b) with different models as described in the text.  \label{q2_rho}}
\end{figure}

A new precision measurement of the reaction $\gamma^*p\rightarrow\rho^0 p$ was published by ZEUS~\cite{zeus_rho}. It was found that the cross section falls steeply with the increasing of $Q^2$ but, unlike it was observed for the $J/\psi$ electroproduction~\cite{zeus_jpsi,h1_jpsi}, it cannot be described by a simple propagator term like $\sigma\propto (Q^2+M^2)^{-n}$, in particular an $n$ value increasing with $Q^2$ appears to be favored. Figure~\ref{q2_rho} reports the cross section for the $\rho^0$ electroproduction versus $Q^2$ compared with several theoretical predictions: the KWM model~\cite{KMW} based on the saturation model, the FSS model~\cite{FSS} with and without saturation and the DF model~\cite{DF}. None of the available models gives a good description of the data over the full kinematic range of the measurement.

\begin{figure}[htbp]
\centering
\includegraphics[width=0.6\textwidth,angle=0]{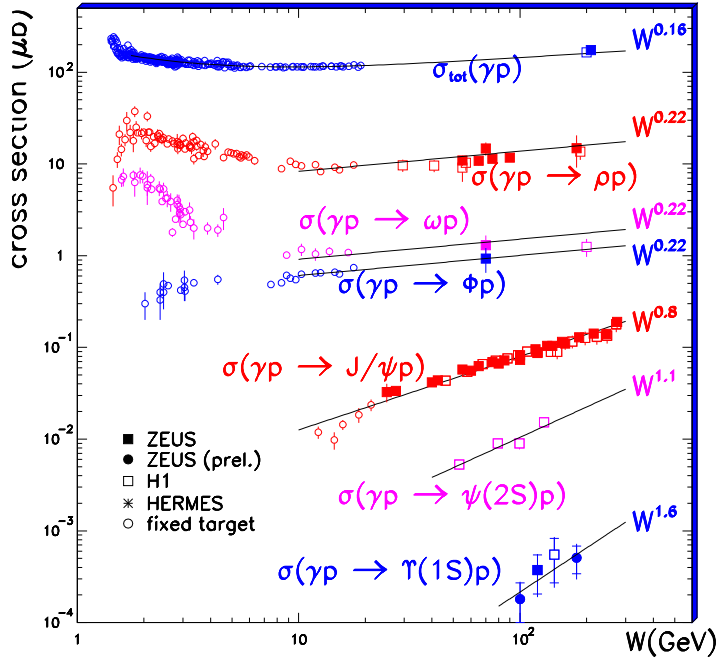}
\caption{The $W$ dependence of the cross section for exclusive VM photoproduction together with the total photoproduction cross section. Lines are the result of a $W^{\delta}$ fit to the data at high $W$-energy values. \label{W_php}}
\end{figure}

The soft to hard transition can be observed looking at the dependence of the VMs photoproduction ($Q^2=0$) cross section from the $\gamma^*p$ centre of mass energy, $W$, where the scale is provided by $M^2$. Figure~\ref{W_php} collects the $\sigma ( \gamma^* p\rightarrow Vp )$ as a function of $W$ from the lightest vector meson, $\rho^0$, to the heaviest, $\Upsilon$, compared to the total cross section. 
The cross section rises with the energy as $W^{\delta}$, where the $\delta$ exponent increases with the hard scale $M^2$ as expected for a transition from the soft to the hard regime. New results on the $\Upsilon$ photoproduction~\cite{upsilon}, recently published by ZEUS, confirmed the steeper rise of $\sigma(W)$ for higher vector meson masses. 

\begin{figure}[htbp]
\centering
\begin{tabular}{cc}
\includegraphics[width=0.5\textwidth,angle=0]{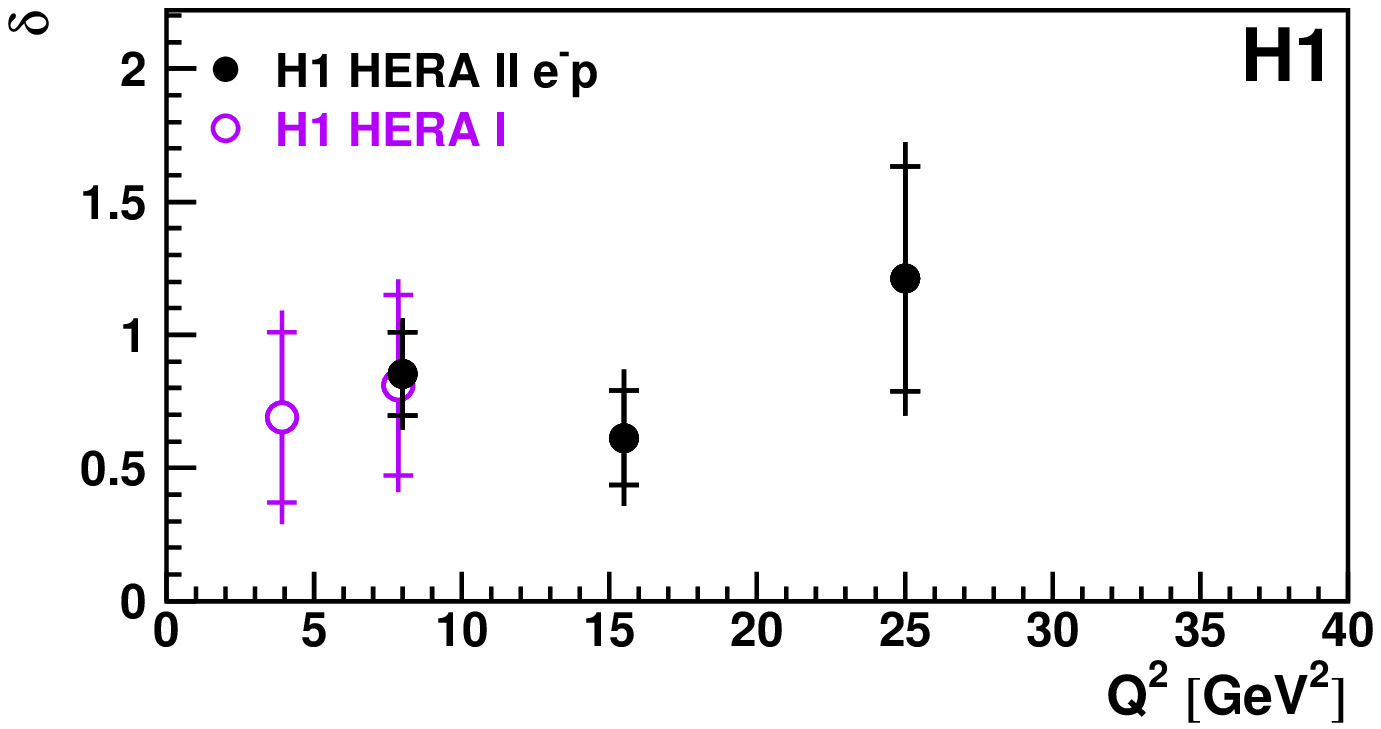}
\includegraphics[width=0.5\textwidth,angle=0]{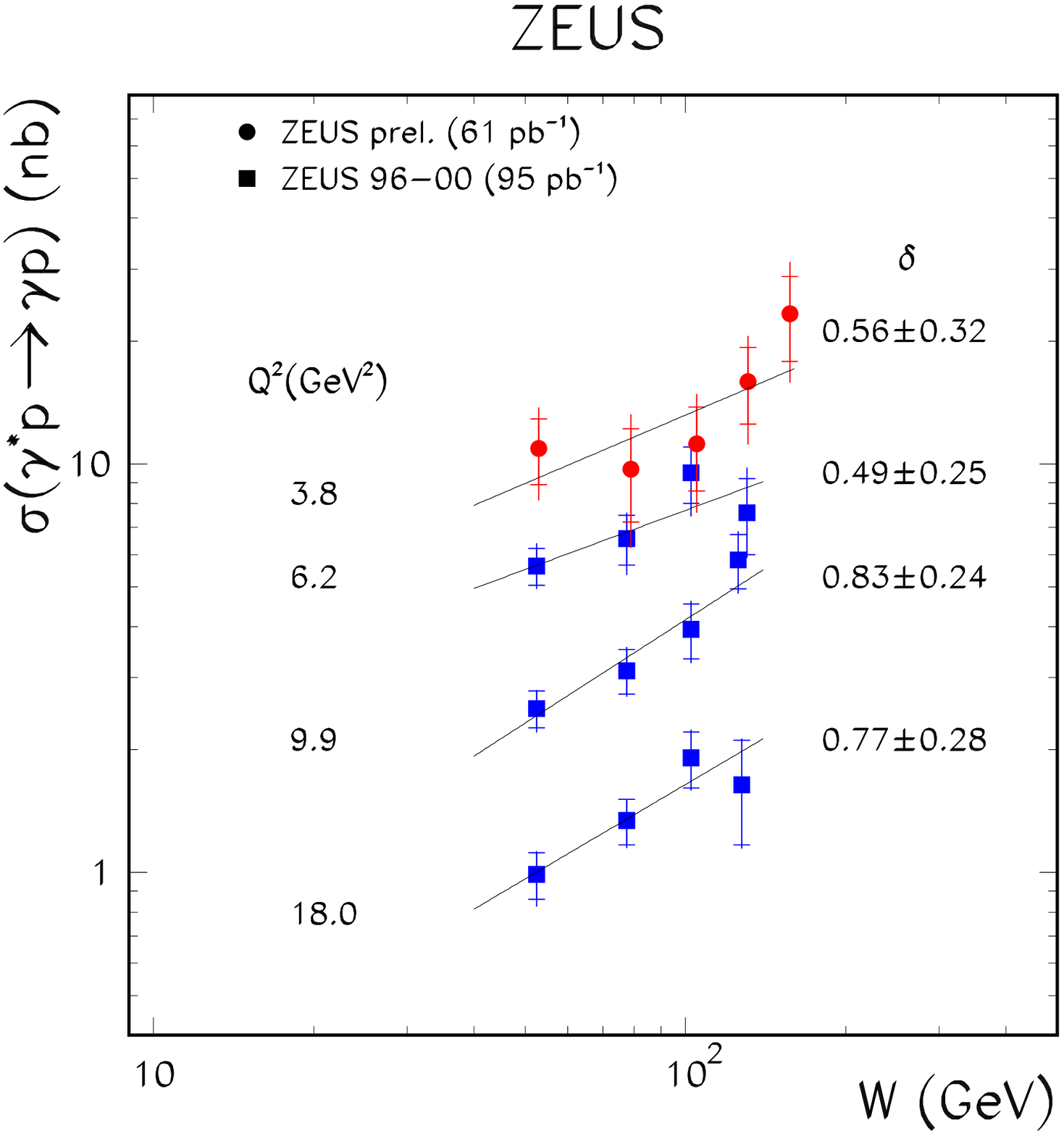}
\end{tabular}  
\caption{The $W$ dependence of the cross section for a DVCS process. Lines come from a $W^{\delta}$ fit to the data. Left: the H1 measurement of the $\delta$ slope as a function of $Q^2$. Right: the new ZEUS preliminary measurement at low $Q^2$ (dots) together with the published measurements (squares). \label{W_dvcs}}
\end{figure}

The transition from the soft to the hard regime can also be studied varying $Q^2$. Recent results were achieved by H1~\cite{h1_dvcs} and ZEUS~\cite{zeus_dvcs} for the exclusive production of a real photon, the Deeply Virtual Compton Scattering (DVCS), where the hard scale is provided only by the photon virtuality, $Q^2$. Figure~\ref{W_dvcs} shows the H1 (left) and the ZEUS (right) results. The steep rise with $W$ of the cross section even at low-$Q^2$, seems to suggest that the most sensitive part to the soft scale comes from the wave function of the pruduced VM. A similar result was obtained for the $J/\psi$ electroproduction~\cite{zeus_jpsi,h1_jpsi}.
 
\begin{figure}[htbp]
\centering
\begin{tabular}{cc}
\includegraphics[width=0.53\textwidth,angle=0]{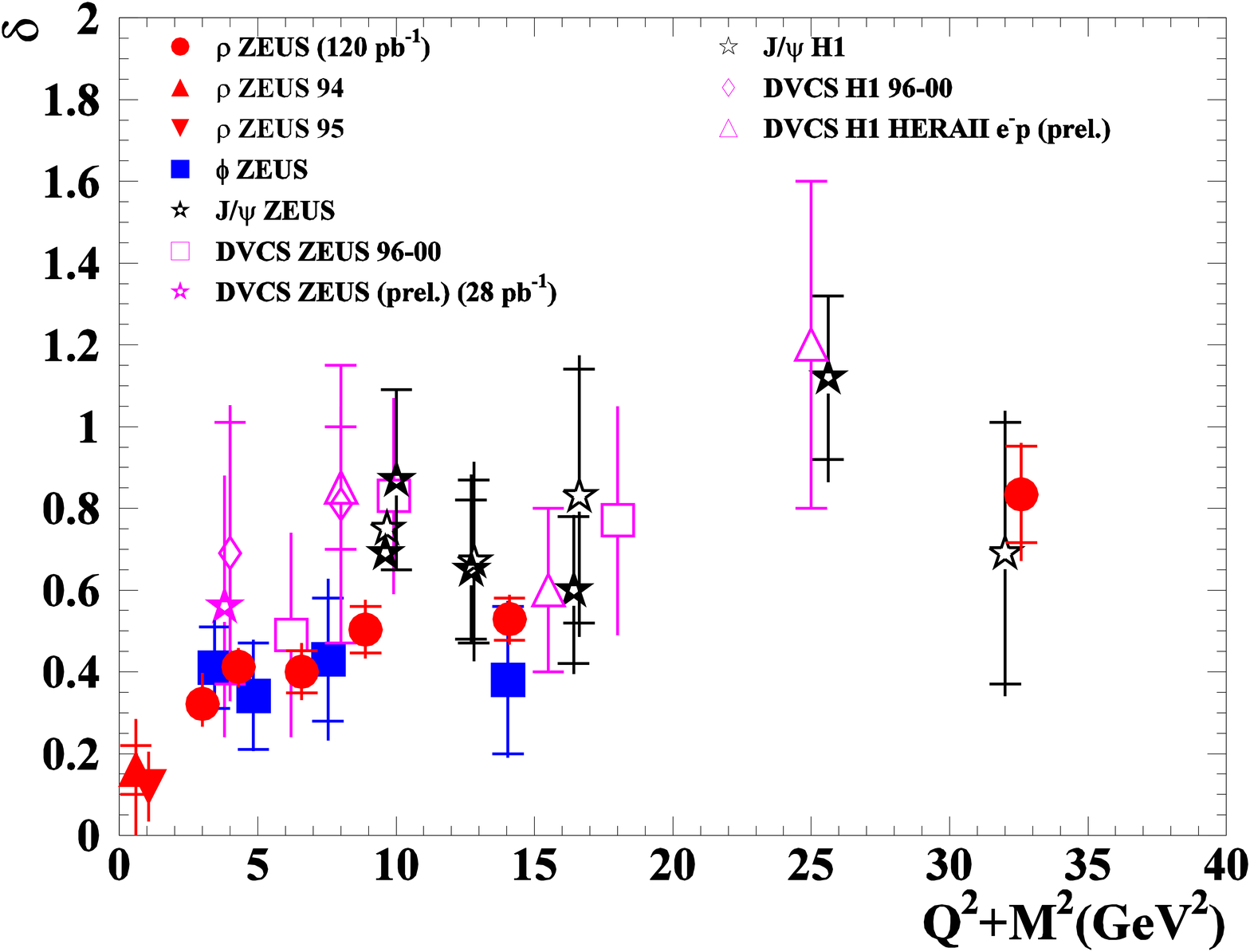}
\includegraphics[width=0.46\textwidth,angle=0]{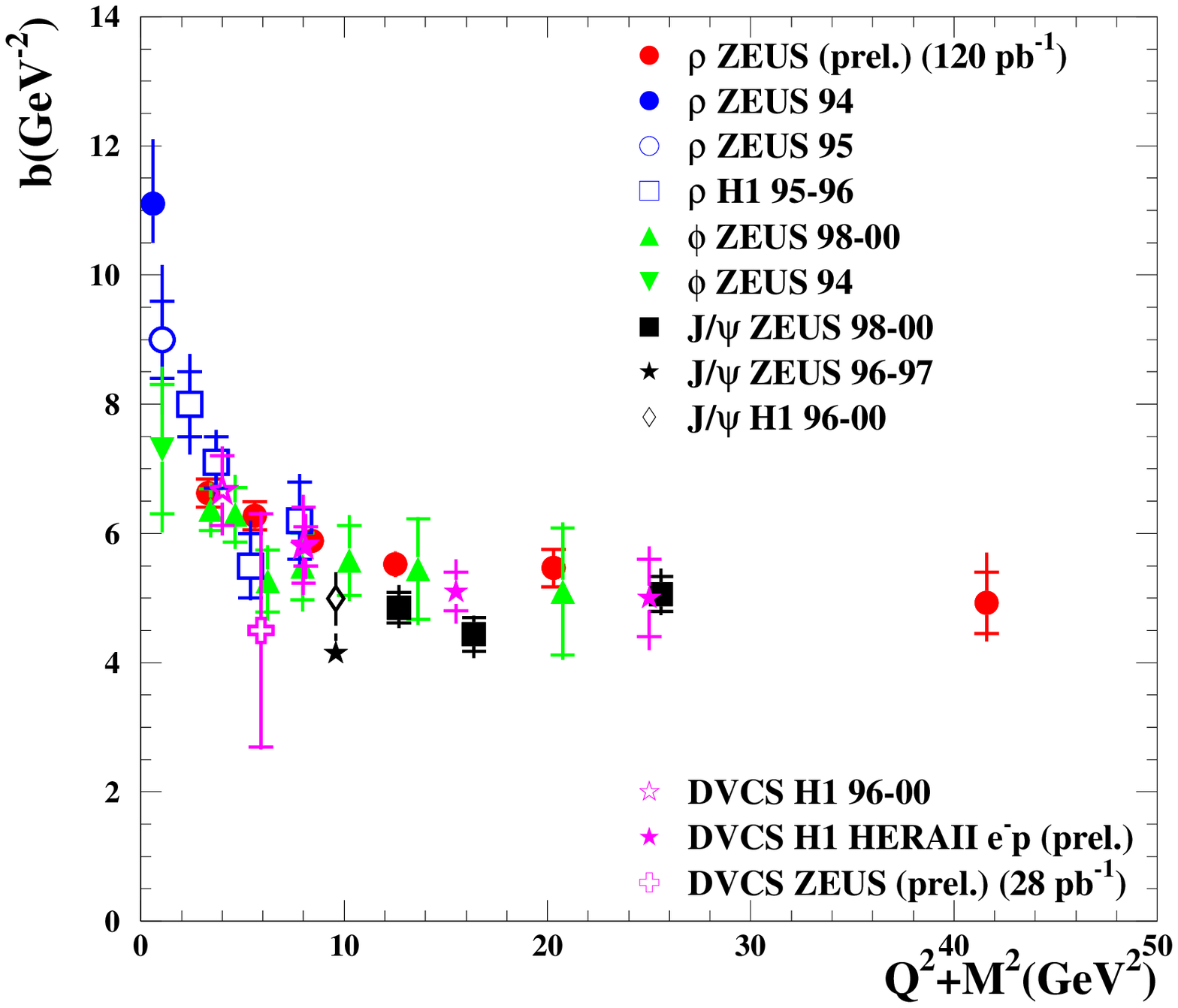}
\end{tabular}   
\caption{The dependence on the hard scale $Q^2+M^2$ of the value $\delta$ (left) extracted from a fit $W^{\delta}$ and of the slope $B$ ($b$ in the figure lable) (right) extracted from a fit $\frac{d\sigma}{dt}\propto e^{B|t|}$ for the exclusive VM electroproduction. DVCS is also included. \label{W_dis}}
\end{figure}

The electroproduction of a large variety of VMs was studied at different $Q^2$ values and the corresponding slope $\delta$ is reported in Fig.~\ref{W_dis} (left) versus the scale $Q^2+M^2$, including the DVCS measurements. Data show a logarithmic shape $\delta\propto \ln(Q^2+M^2)$ and the behaviour seems to be universal with $\delta$ increasing from 0.2 at low scale, as expected from a soft Pomeron exchange~\cite{Wsoft} to $\sim 0.8$ at large scale values. 

\section{$t$ dependence of the cross section}

The differential cross section as a function of $t$, the four-momentum tranfer at the proton vertex, can be parametrised by an exponential fit: $\frac{d\sigma}{d|t|}\propto e^{B|t|}$. Figure~\ref{W_dis} (right) reports the collection of the $B$ values versus the scale $Q^2+M^2$ for the electroproduction of VMs and DVCS, with $B$ decreasing from $\sim 11\; GeV^{-2}$ to $\sim 5\; GeV^{-2}$ as expected in hard regime.  

\begin{figure}[htbp]
\centering
\begin{tabular}{cc}
\includegraphics[width=0.5\textwidth,angle=0]{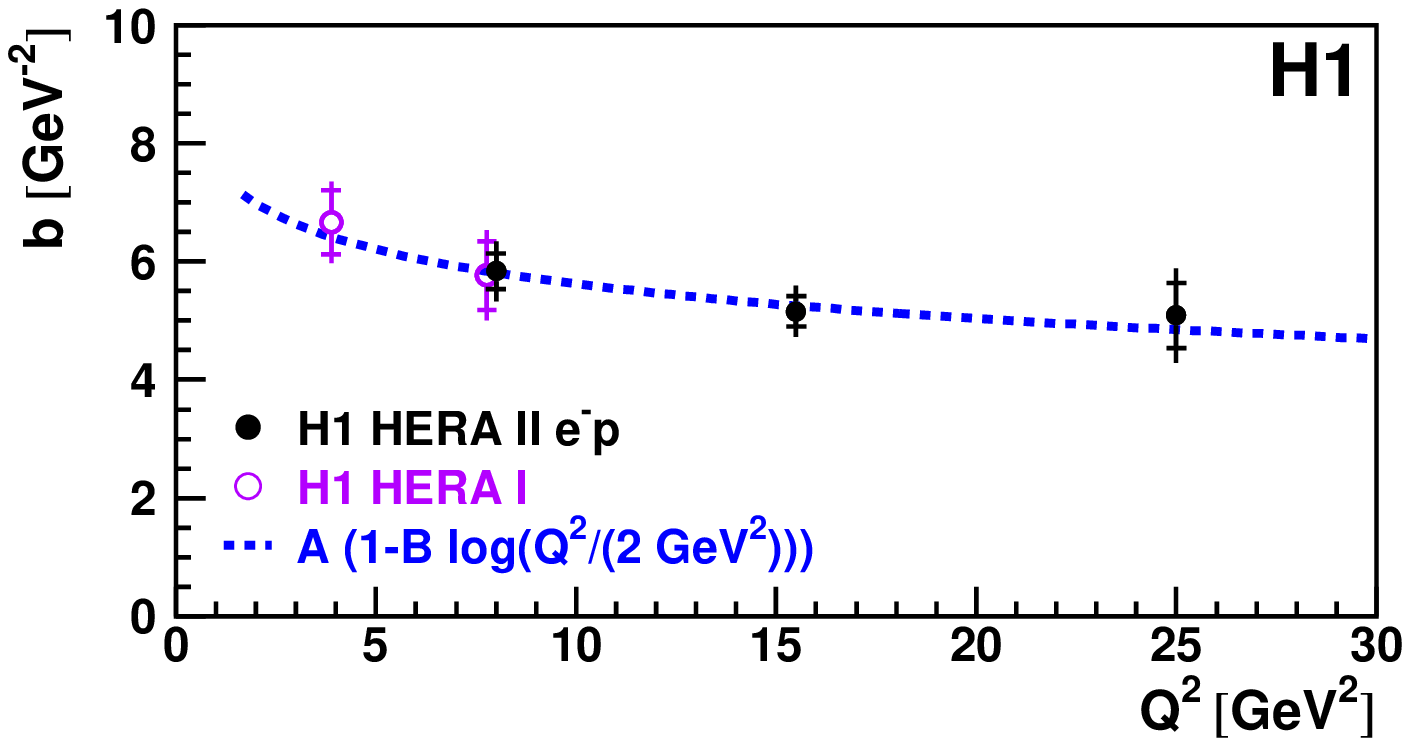}
\includegraphics[width=0.5\textwidth,angle=0]{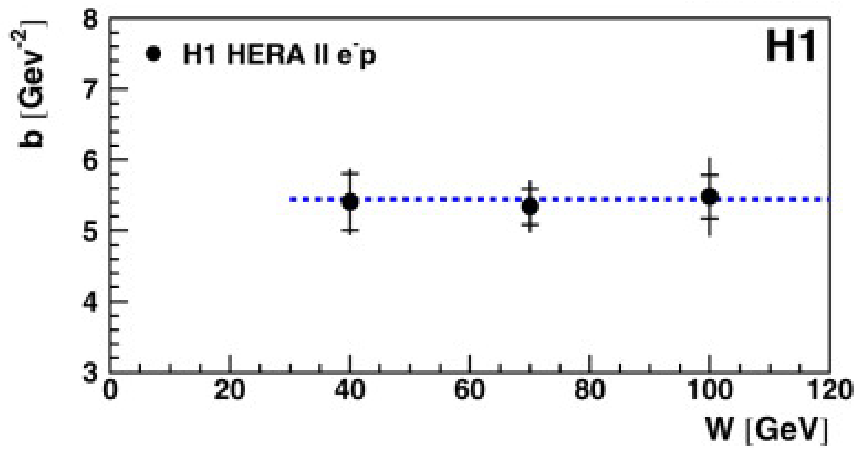}
\end{tabular}
\caption{The $t$ slope parameter $B$ ($b$ in the figure lable) as a function of $Q^2$ (left) and $W$ (right).}
\label{b_dvcs}
\end{figure}

The measurement of $d\sigma/d|t|$ for the DVCS process, recentrly published by the H1 Collab~\cite{h1_dvcs}, where $t$ was obtained from the transverse momentum distribution of the photon, studied $B$ versus $Q^2$ and $W$ as shown in Fig.~\ref{b_dvcs}. $B$ seems to decrease with $Q^2$ up to the value expected for a hard process but it doesn't depend on $W$.   
A new preliminary ZEUS measurement~\cite{zeus_dvcs} of $d\sigma/d|t|$ has been achieved from a direct measurement of the proton final state of using a spectrometer based on the roman pot thechnique. The result $B=4.4\pm 1.3~(stat.)\pm 0.4~(syst.)~GeV^{-2}$, measured at $Q^2=5.2~GeV^2$ and $W=104~GeV$, is consistent, within the large uncertainties due to the low acceptance of the spectrometer, with the H1 result~\cite{h1_dvcs} of $B=5.45\pm 0.19~(stat.)\pm 0.34~(syst.)~GeV^{-2}$ at $Q^2=8~GeV^2$ and $W=82~GeV$.

Since the $B$ value can be related via a Fourier transform to the impact parameter and assuming that the exclusive process in the hard regime is dominated by gluons, the relation $\langle r^2\rangle=2b(\hbar c)^2$ can be used to obtain the radius of the gluon confinement area in the proton. $b\sim 5\;GeV^2$ corresponds to $\langle r^2\rangle\sim 0.6\;fm$ smaller than the proton radius ($\sim 0.8\;fm$) indicating that the gluons are well contained within the charge-radius of the proton.

\section{The Pomeron in DVCS at HERA}\label{s1}
Exclusive production of vector states (real photon, VMs, lepton pairs) via deeply virtual scattering, $ep\rightarrow epV$, is interesting for many reasons, above all as a source in extracting informations about General Parton Distributions (GPDs). Of particular interest is DVCS in which the outgoing real photon, interferring with the photon coming from bremsstrahlung (Bethe-Heitler process), offers a holografic picture of the nucleon.

A $Q^2$-dependence of generalized parton distributions (GPD) can be found from QCD evolution, similar, although less explored than that of ordinary parton distribution, obeying the DGLAP evolution equation \footnote{for a discussion of GPD evolution see the Appendix}.

In the present paper, we analize and test an explicite model for DVCS. The model can be used to infer the GPD by a relevant deconvolution procedure.

After the shut-down of the electron-proton collider HERA,
the H1 and ZEUS collaborations have
left a huge heritage of data still waiting for a better
understanding. In particular, it concerns the properties of the
Pomeron trajectory as seen in VM electroproduction, $ep\rightarrow e V p$, and in DVCS, $ep\rightarrow e\gamma p$.
There are many papers discussing in details the form and the values of the
parameters of the Pomeron trajectory as well as their possible
$Q^2$ dependence (for recent reviews see, e.g. \cite{Levy}).
In general they  introduce a linear Regge-trajectory of the Pomeron
\begin{equation}
    \alpha_P(t,Q^2)=\alpha_0(Q^2)+\alpha '(Q^2)t,
\label{trlin}
\end{equation}
where $t$ is the squared four-momentum transferred at the proton vertex and $Q^2$ is the virtuality of the exchanged photon.
For the \textquotedblleft effective\textquotedblright~Pomeron they uses the standard Regge pole parametrization of the scattering amplitude
\begin{equation}
A(s,t,Q^2)=A_0e^{B(t,Q^2)}\Biggl({s\over{1GeV^2}}\Biggr)^{\alpha_P(t,Q^2)},
\end{equation}
where $s=W^2$ is the squared $\gamma^* p$ centre-of-mass energy and $B(t,Q^2)$ is related to the radius associated with the proton vertex.


The problem of constructing Regge-type models for currents
(lepton-hadron scattering) or the off-mass-shall continuation of
the analytic $S$-matrix has already a long history,
still lacking a consistent solution. The concepts of the analytic
$S$-matrix formally may even be incompatible with, or inapplicable
to processes like Compton scattering. Nevertheless, the need for a
theoretical framework to describe high-energy lepton-hadron
scattering served as an \textquotedblleft excuse\textquotedblright~in using $Q^2-$dependent
Regge-type models for small $x$ (large $s\sim Q^2/x)$ processes,
like Compton scattering. Moreover, to meet the apparent
acceleration of the rise of the structure functions with $1/x$
towards larger $Q^2$, a $Q^2$ dependent Pomeron was introduced,
in complete discords with Regge factorization, in which the
dependence on the mass or virtuality of the external particle
(photon or meson in our case) can enter at most through the
relevant vertex function.  This inconsistency was circumvented by
calling this object \textquotedblleft effective Pomeron\textquotedblright~(or, more generally, a Reggeon),
implying that it accommodates by more than a single Regge
exchange, without specifying the content. Very often, however,
this was ignored and the parameters of the Pomeron have been
extracted from simple formula like in Eq.~\ref{trlin}. Most of the conclusions,
especially the appearance a large and ever increasing intercept (hard Pomeron)
and its small slope (flat Pomeron) were used to confirm perturbative QCD.

At the same time, it was shown~\cite{JKLP} that the inclusion of a
sub-leading Regge contribution modifies the parameters of the
Pomeron, in particular the observed rise of the structure function
partly is due to the decrease of the sub-leading contribution, and
the $Q^2$ dependent residue functions can provide for the apparent
\textquotedblleft hardening\textquotedblright~of the Pomeron.

Recently a number of Regge/Pomeron-type models for DVCS appeared
in the literature~\cite{CFFJP,Mullerfit,Szczepaniak,Mueller}.

In the present paper, we analyze the high-energy data on DVCS
collected by the H1 and ZEUS
detectors at HERA~\cite{H1_08,zeus_dvcs,H1,ZEUS}. To do so, we use a model
for the Pomeron elaborated and fitted to the data in Ref.~\cite{CFFJP}. 

In most of the Regge/pole models used in analyzing the experimental data, a linear Pomeron trajectory is used. Although linear trajectories contradict to the postural of the $S$-matrix (analyticity and unitarity), to perturbative QCD calculations (BFKL)~\cite{BFKL}, as well as disagree with precize fits to the data, they can serve as a simple approximation to the observed phenomena at small $|t|$~\cite{CFFJP} however the linear behavior is replaced by a logarithmic one~\footnote{For further arguments see the Ref.~\cite{CFFJP}}.  

For this reason we use a simple Regge pole model for the
Pomeron, however, contrary to most of the known models,
our Pomeron trajectory is essentially nonlinear.

The scattering amplitude has the form

\begin{equation}\label{A2}
A(s,t,Q^2)_{\gamma^* p\rightarrow\gamma p}=
-A_0e^{b\alpha(t)}e^{b \beta(z)}(-is/s_0)^{\alpha(t)}=
-A_0e^{(b+L)\alpha(t)+b\beta(z)},
\end{equation}
where $L\equiv\ln(-is/s_0)$. The trajectory at the $pIPp$ vertex is
\begin{equation}\label{alpha}
\alpha(t)=\alpha_0-\alpha_1\ln(1-\alpha_2 t).
\end{equation}

whereas the trajectory at the $\gamma^*IP\gamma$ vertex is
\begin{equation}\label{beta}
\beta(z)=\alpha_0-\alpha_1\ln(1-\alpha_2 z),
\end{equation}
with $z=t-Q^2$ - a new variable introduced in Ref.~\cite{CFFJP}.
Notice that the presence of the \textquotedblleft minus\textquotedblright
sign in Eq.~\ref{A2} is
important for the linear forms, e.g. $ImA(t=0)$, proportianal to the
the total cross section, or for the ratio $\rho=ReA/ImA$, but
it is irrelevant for the squared modulus we are interested in.

Similar to~\cite{Mullerfit}, we consider only the helicity
conserving amplitude. For not too large $Q^2$ the contribution
from longitudinal photons is small (it vanishes for $Q^2=0$).
Moreover, at the high energies typical of the HERA collider, the
amplitude is dominated by the helicity conserving Pomeron exchange
and, since the final photon is real and transverse, the initial
one is also transverse and the helicity is conserved.
The electroproduction of
vector mesons requires the account for both the longitudinal
and transverse cross sections.

Generally speaking, any DVCS reaction admits the contribution from a
subleading Regge exchanges, most important of which is the $f$
trajectory, made of a quark and an antiquark, not gluons
like the Pomeron. It differs only by the values of the Regge-trajectory parameters (lower intercept and steeper slope).

The cross section can be calculated as

\begin{equation}{d\sigma\over{dt}}(s,t,Q^2)={\pi\over{s^2}}|A(s,t,Q^2)|^2.
\label{A6}
\end{equation}

with the slope of the $d\sigma/d|t|$ cross section coming from
\begin{equation}B={d\over{dt}}\ln\Bigl({d\sigma\over{dt}}\Bigr).
\label{A7}
\end{equation}

\begin{figure}[h]
\begin{center}
\hspace{-1.cm}
\includegraphics[width=0.8\textwidth,angle=0]{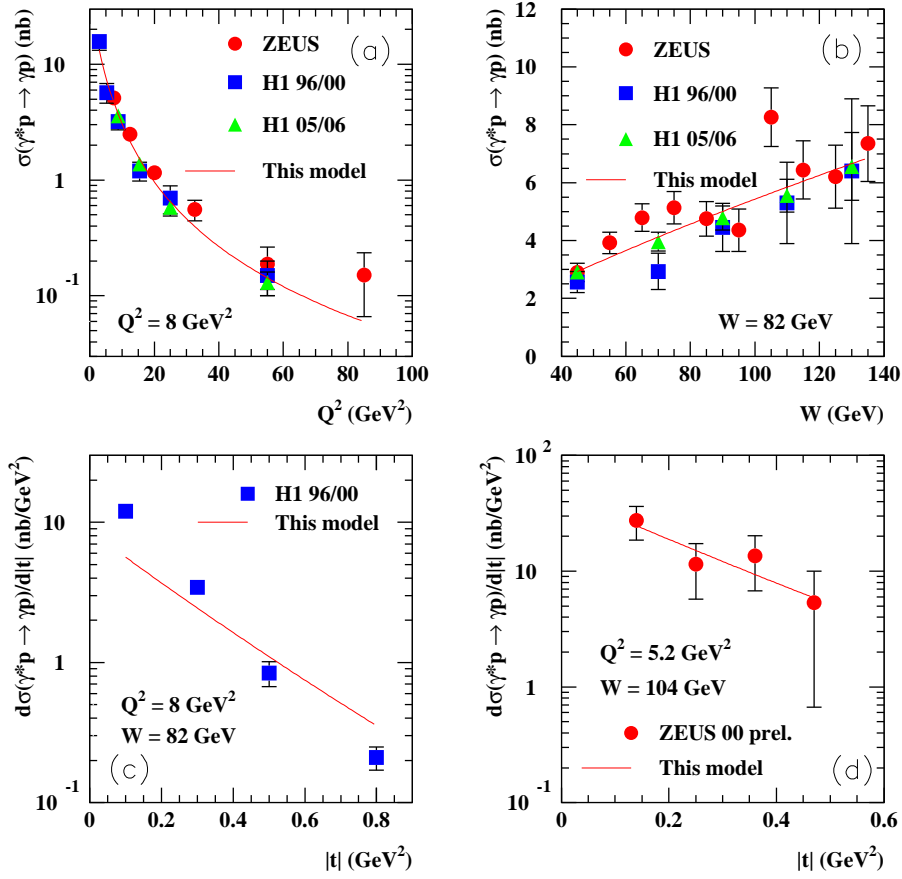}
\caption{\small \it {Upper panel: Energy (a) and $Q^2$ (b) dependence of the DVCS cross section, the model is fitted to the HERA data. Lower panel: the prediction for the $t$ dependence of the cross section is compared with H1 (c) and ZEUS (d) data with only the normalisation kept as a free fit parameter. Error bars include both statistical and systematic uncertainties summed in quadrature. The error bars include both the statistical and systematic uncertainties summed in quadrature.}} \label{fig:fit_new}
\end{center}
\end{figure}

\begin{figure}[h]
\begin{center}
\hspace{-1.cm}
\includegraphics[clip,scale=0.5]{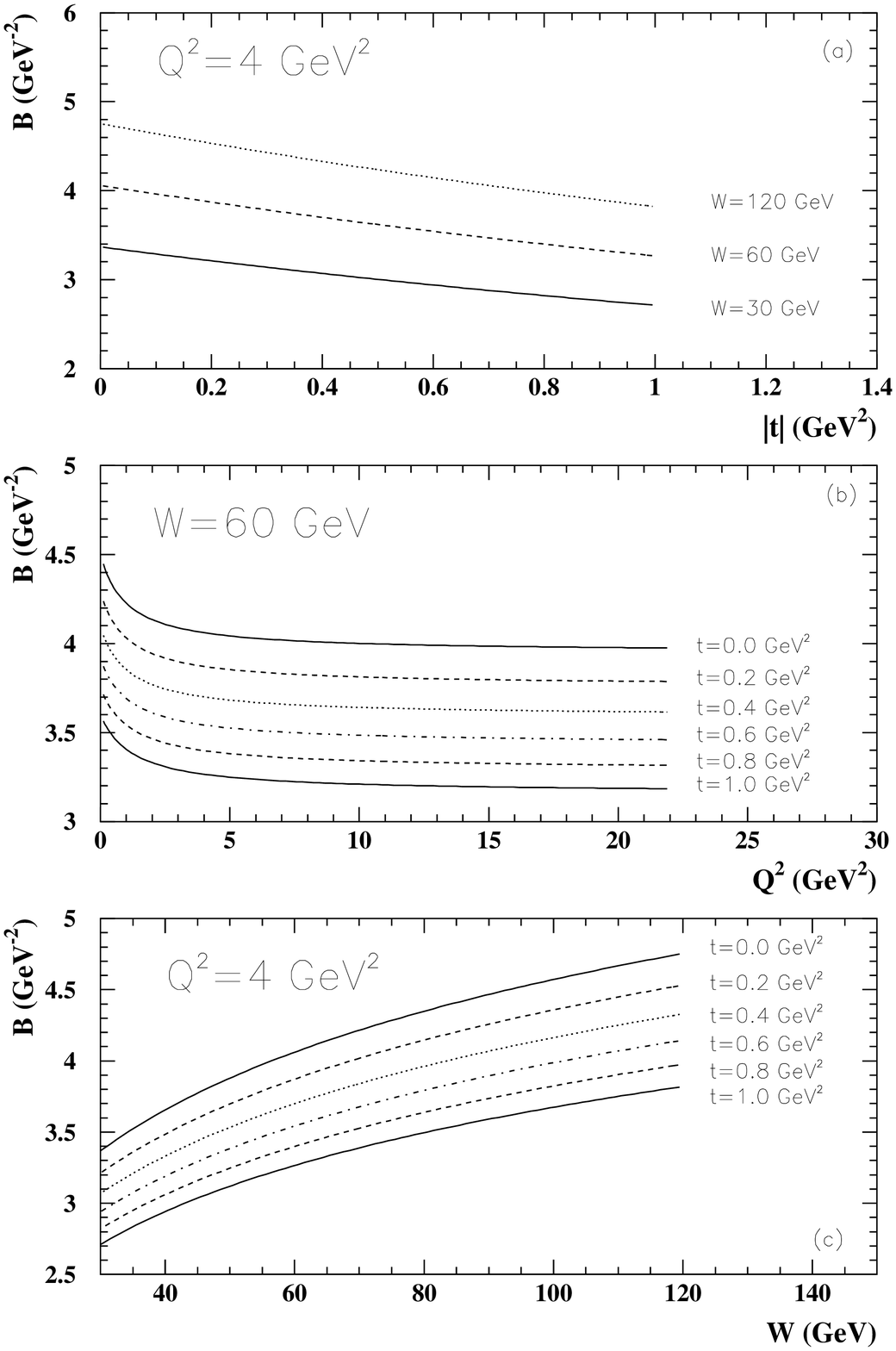}
\caption{\small \it {Slope of the cross section as calculated from Eq.~\ref{A6}. for several values of $W$, $Q^2$ and $t$}.} \label{fig:slope}
\end{center}
\end{figure}

The model contains quite a number of parameters but some of them ($s_0$, $\alpha_1$, $\alpha_2$) can be fixed by theoretical constraints, as explained in Ref.~\cite{CFFJP}. Fits to the DVCS data collected at HERA~\cite{H1,H1_08,ZEUS} were performed and the results are shown in Fig.~\ref{fig:slope}, with the values of the fit parameters, $A_0$, $b$, $\alpha_0$ quoted in Table~\ref{fig:slope}.
 Note that $\sigma(W)$ is sensitive to the Pomeron intercept, $\alpha_0$, and the corresponding fit was performed by keeping $b_P=b=1.0$ fixed, while the $\sigma(Q^2)$ is sensitive to $b$ and the corresponding fit was done at fixed $\alpha_0=1.2$.
Figures~\ref{fig:fit_new}a,b show that the model can
reproduce all the features of data versus $Q^2$ and $W=\sqrt{s}$.
Results suggest that the Pomeron in DVCS has an intercept, $\alpha_0\simeq 1.2$ graeter than that in hadronic reactions, but a slope, $\alpha '\simeq 0.25$, typical of hadronic scattering and not so flat as it was observed for the diffractive electroproduction of vector mesons ($\alpha '\sim 0.1\simeq 0.5 \alpha_hh '$) in hard regime.

\begin{table} [htbp]
\begin{center}
\begin{tabular}{|l|c|c|c|c|}
\hline
Parameter & $\sigma_{DVCS}$ vs $Q^2$ & $\sigma_{DVCS}$ vs $W$ &  $d\sigma/d|t|$ H1 & $d\sigma/d|t|$ ZEUS\\
\hline
$A_0$   & 0.13 $\pm$ 0.01             & 0.10 $\pm$ 0.01 & 0.17 $\pm$ 0.01 & 0.44 $\pm$ 0.10 \\
$\alpha_0$     & fixed at 1.2             & 1.23 $\pm$ 0.006  & fixed at 1.2 & fixed at 1.2 \\
$b$       & 0.96 $\pm$ 0.07             & fixed at 1.0   & fixed at 1.0 & fixed at 1.0 \\
\hline
$\tilde{\chi}^2$ & 1.39                  & 1.21         & 16.8  & 0.3 \\
\hline
\end{tabular}
\end{center}
\caption{\small \it {Values of fitted parameters and the corrispondent $\tilde{\chi}^2$ value.}}
\label{table}
\end{table}

The fit of $d\sigma/d|t|$ was performed with all the parameters fixed excepted for the normalisation. The model does not agree with the H1 measurements (see Fig.~\ref{fig:fit_new}c) but it is compatible with the new ZEUS preliminary results (see Fig.~\ref{fig:fit_new}d). The $t$ variable was calculated in H1 by the transverse four-momenta of the scattered electron and the real photon, using the apprroximation $t\simeq|P_{T_e}+P_{T_\gamma}|^2$, while in ZEUS  a particular silicon microstrips spectrometer, based on the roman pots technique, was used in order to have a direct measurement of the scattered proton momentum $\vec{p}\:'$ and then $t$ being calculated from the quantity: $x_L=\frac{p_{z}'}{p_{beam}}$, by the phormula: $t=\frac{p'^{2}_{x}+p'^{2}_{y}}{x_L}+E^{2}_{beam}\frac{(x_{L}-1)^2}{x_L}$. The really low acceptance of this spectrometer is the reason of the poor statistics of the ZEUS data, howewer it offers a really pure selection of diffractive events, not affected by any not-diffractive background. The agreement with the ZEUS preliminary measurements encourages us to be still confident in our predictions, till they will be checked by the new HERA data analyses (now in progress with and without roman pots spectrometer).

\begin{figure}[h]
\begin{center}
\hspace{-1.cm}
\includegraphics[clip,scale=0.5]{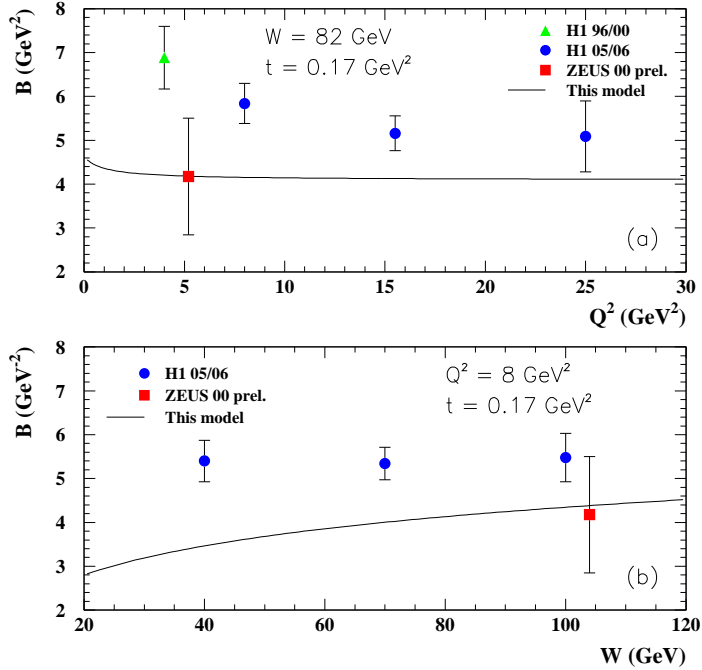}
\caption{\small \it {Slope of the cross section vs $Q^2$ and
$W$ as calculated from Eq.~\ref{A7}, compared with the HERA data.}} \label{fig:slope_q2}
\end{center}
\end{figure}

The slope of $d\sigma/d|t|$, calculated according to Eq.~\ref{A7}, is depicted in Fig.~\ref{fig:slope_q2} as a function of $Q^2$ and $W$ and is compared with de HERA measurements. The local slope is predicted to slightly rise with $W$ but to be almost independent on $Q^2$.

Eq.~\ref{A7} was then used to make a collection of figures showing the slope dependence for different values of $W$ and $t$.
The $t$ dependence of the local slope calculated from Eq.~\ref{A7} is shown in Fig.~\ref{fig:slope}a at fixed $Q^2=4\;GeV^2$ for three different values of energy.


Figure~\ref{fig:slope}b depicts the $Q^2$ dependence of the local slope at $W=60\;GeV$ for several values of $t$.

Figure~\ref{fig:slope}c shows the enerrgy dependence of the local slope at $Q^2=4\;GeV^2$ for several values of $t$.




\section{Conclusions}

\begin{itemize}
    \item After the shut-down of the HERA Collider, the ZEUS and H1 experiments have left a huge heritage of experimental results on the diffractive production of vector mesons and real photons still waiting for a better understanding. Nevertheless, new analyses on HERA data are still in rpogress.
    \item We show that data on DVCS from HERA can be fitted with
a semi-hard Pomeron, not being flat.
    \item The Pomeron trajectory could be a non-linear function; it is
nearly linear at small and moderate values of $|t|$, leveling off
at large $|t|$. The same logarithmic trajectory could be used in hadronic
reactions, for example in analyzing, with higher precision, the
future diffractive exclusive data from LHC, where details, such as the two-pion
threshold \cite{C_T}, neglected here, should be taken into
account.
    \item Here we used a simple \textquotedblleft minimal\textquotedblright~model, with a small number of the free parameters, which, in view of the limited
number of data points, made possible the convergence of the
minimization procedure. A more detailed analyses should include
also the individual contributions from the longitudinal and
transverse cross sections, though their ratio $R$, spin effects,
reaction ratios and comparison with possible QCD predictions,
especially concerning the evolution in $Q^2$ (see next item and
Appendix).
    \item A complete solution for the $Q^2$ evolution in DVCS has
not been yet found, although partial results are known from the
literature. In a simplified, pragmatic approach
the $Q^2$ and $t$ dependence in a DVCS (actually not
factorizable!), reduces the problem to the evolution of ordinary
DIS by the DGLAP procedure. These attempts are summarized in
Appendix.
    \item The model and the fitting procedure presented in this
paper can be extended to high-energy vector meson production. We
intend to do so in a forthcoming publication. Let us notice also
that any consistent extension/interpolation of the Regge approach
to low energies (resonance region) should involve dual models, as
it was shown~\cite{JFMPP} for the Pomeron component in case of
$J/\Psi$ scattering.
\end{itemize}

\medskip

{\bf Acknowledgment} 
We would like to thank R.~Fiore, F.~Paccanoni and V. Nikitin for the helpful discussions.
L.~Jenkovszky thanks the University of Calabria, where part of this work
was done, for its hospitality and support. 

\newpage
\bigskip
\section*{APPENDIX. GPD and QCD evolution of DVCS}
From Eq.~\ref{A2} we get the real and imaginary parts of the DVCS
amplitude at $Q^2=Q_0^2,$ that will be useful for the evolution:
\begin{equation}
Im A_{DVCS}(x_{Bj},t,Q_0^2)=
sin[\pi\alpha(t)/2]G(t,Q_0^2)\Bigl(\frac{Q_0^2}{s_0x_{Bj}}\Bigr)^{\alpha(t)},
\label{ImA}
\end{equation}

\begin{equation}
Re A_{DVCS}(x_{Bj},t,Q_0^2)=
-cos[\pi\alpha(t)/2]G(t,Q_0^2)\Bigl(\frac{Q_0^2}{s_0x_{Bj}}\Bigr)^{\alpha(t)},
\label{ReA}
\end{equation}
where
\begin{equation}
G(t)=e^{b[\alpha(t)+\beta(t,Q^2_0)]}.
\end{equation}
Skewendess is defined in terms of the Bjorken variable $x_{Bj}$ as
$\xi\simeq 2x_{Bj}-x_{Bj}$ and v.v., $x_{Bj}\simeq 2\xi/(1+\xi).$

The aim is to evolve this amplitude to higher values of $Q^2,\ \
Q^2>Q_0^2\sim 1 Gev^2$ and, at the same time, to grasp the most
important properties of the corresponding generalized parton
distributions (GPD).

The singlet combination of GPD, corresponding to the exchange in
the $t-$ channel of a $C=+1$ charge conjugation, even signature
reggeon is~\cite{Diehl}
\begin{equation} H^{q(+)}(x,\xi,t)=H^q(x,\xi,t)-H^q(-x,\xi,t),
\end{equation}
with $H^{q(+)}(x,0,0)=q(x)+\bar q(x),$ where $q(x)$ is the usual
quark distribution for quark $q.$ Its evolution is well known. The
\textquotedblleft valence\textquotedblright combination, odd signature,
is even under the change: $x\rightarrow -x.$

The point $x=\xi,$ where DGLAP $(|x|>\xi)$~\cite{DGLAP} and ERBL
$(|x|<\xi)$~\cite{ERLB} regimes meet, is directly accessible to to
experiment since, at leading order in $\alpha_s,$ the imaginary
amplitude part of the amplitude $A$ for DVCS is proportional to a
GPD at $x=\xi$~\cite{FJM}
\begin{equation}
Im A_{DVCS}(x_{Bj},t,Q_0^2)=-\pi\sum_q
e_q^2H^{q(+)}(\xi,\xi,t,Q_0^2),
\end{equation}
where $H^{q(+)}(\xi, \xi, t, Q_0^2)$ is known from Eq.~\ref{ImA}.
Unfortunately, the evolution equation for $H^{q(+)}(\xi, \xi, t,
Q_0^2)$ involves $H^{q(+)}(x, \xi, t, Q_0^2)$ for $x$ in the
interval $[0,1]$ (for DVCS), to be known.

An approximate solution based on conformal invariance, valid at
one loop level is possible~\cite{Bukhvostov}. The off-forward
matrix elements of a set of twist-two operators, whose leading-log
evolution is diagonal due to conformal symmetry, are the
Gegenbauer polynomials. In the forward kinematics, the Gegenbauer
polynomials reduce to polynomials in $x_{Bj}.$ In Ref.~\cite{Shuvaev}
an explicit transformation has been constructed
relating the off-forward and the forward evolutions.

Usually it is very difficult to apply the transformation
constructed in Ref.~\cite{Shuvaev} to a realistic problem but, in
our case, the simplicity of the input at $Q_0^2$ makes possible an
explicit asymptotic solution for small $\xi_{Bj}$ and $\xi\sim
x_{Bj}/2.$ It has been shown in Ref.~\cite{Shuvaev1} that, in the
phenomenologically important small-$\xi$ and $|t|$ region, the
off-diagonal distributions are determined unambiguously by the
small-$x$ behavior of the diagonal parton distributions. By
setting \begin{equation} H(x,\xi)\equiv H(x,\xi,t,\mu^2),\ \
(-1\leq x_{Bj}\leq 1), \end{equation} where the values of $t$ and
$\xi$ do not change in the evolution to a higher scale $\mu^2$, it
was shown in Ref.~\cite{Shuvaev1} that at small $\xi$
\begin{equation}
H^q(x,\xi)=\int_{-1}^1dx'\Biggl[{2\over {\pi}}\Im
\int_0^1{ds\over{y(s)\sqrt{1-y(s)x'}}}\Biggr]{d\over{dx'}}\Biggl({q(x')\over{|x'|}}\Biggr),
\label{Hq}
\end{equation}
\begin{equation}
y(s)={4s(1-s)\over{x+\xi(1-2s}}.
\end{equation}

The proof of Eq.~\ref{Hq} uses the properties of the diagonal parton distributions
$$q(x'\rightarrow 1)\rightarrow 0,\ \ q^s(x')=-q^s(-x'),\ \
q^{ns}(x')=q^{ns}(-x'),$$ that corresponds to the symmetry
relations
$$H^i(x,\xi)= H^i(x,-\xi),\ \ (i=(q,w), (q,ns),$$
$$H^{q,s}(x,\xi)= H^{q,s}(-x,\xi),$$
$$H^{q,ns}(x,\xi)= H^{q,ns}(-x,\xi),$$
where, as usual, $s$ stays for \textquotedblleft singlet\textquotedblright quark contribution and
$n_s$ means \textquotedblleft non-singlet\textquotedblright. At small $\xi$, the same anomalous
dimensions $\gamma_N$ control both the diagonal and off-diagonal evolution.

Consider now the prediction of \cite{Shuvaev1} for the
off-diagonal distribution $H^q(x,\xi)$ by making the physically
reasonable small-$x$ assumption that \begin{equation}
xq(x)=N_qx^{-\lambda_q}, \end{equation} where  $\lambda_q$ may
depend also on $t,\ \ \lambda_q=\lambda_q(t).$ Then eq. (18) can
be integrated
\begin{equation}
H^q(x,\xi)=N_q{\Gamma(\lambda+5/2)\over{\Gamma(\lambda+2)}}{2\over{\sqrt{\pi}}}
\int_0^1ds\Biggl[{4s(1-s)\over{x+\xi(1-2s)}}\Biggr]^{\lambda_q+1},
\label{Hqbis}
\end{equation}
where, for singlet quarks (i.e. when $\lambda_q<1$) the integral
is a principal value integral and Eq.~\ref{Hqbis} becomes integrable for
any $\lambda_q<1.$

The distribution $H^q(x,\xi),$ for small $\xi,$ has the form~\cite{Shuvaev1}
$$H^q(x,\xi)=\xi^{-\lambda_q-1}F_q(x/\xi)$$
with
$$H^q(\xi,xi)=\xi^{-\lambda_q-1}F_q(1),$$
where $F_q(z)$ is a function determined by Eq.~\ref{Hqbis}. Similarly, if
$xg(x)=N_gx^{-\lambda_q},$ then
$H^q(\xi,\xi)=\xi^{-\lambda_g}F_g(1)$ at small $\xi.$  Viceversa,
since the amplitude in Eqs.~\ref{ImA}, and~\ref{ReA} is of Regge form and is
proportional to $\xi^{-\alpha(t)}$ for small $\xi$ and fixed
$Q^2,$ then
\begin{equation}
\sum_qe^2_qH(\xi,\xi)=-{1\over\pi}Im
A_{DVCS}\propto\xi^{-\alpha(t)}
\label{sume}
\end{equation}
and the parton distribution that appears in Eq.~\ref{Hq} has the form
\begin{equation}
xq(x)\propto x^{-\alpha(t)+1},
\end{equation}
where $xq(x)$ can represent $F_2(x)$ if we start from Eq.~\ref{sume}.

An approximate analytic evolution of the input singlet parton
distribution is well known if the input is of the Regge-like form. Since only the diagonal parton distributions are
involved in the calculations, one can resort to libraries that
generate parton distributions but, since we are interested in the
small-$\xi$ region. Equation~\ref{Hq} then determines the $Q^2$
dependence of $H^q(x,\xi)$ and, finally, Eq.~\ref{sume} gives the scale
dependence of $Im A_{DVCS}$. The knowledge of $H^{q(+)}(x,\xi)$
fixes also, at leading order in $\alpha,$ since
\begin{equation}
A_{DVCS}(\xi, t, Q^2)=\sum_qe^2_q\int_0^1H^{q(+)}(x,\xi, t,
Q^2)\Biggl({1\over{x-\xi+i0}}+{1\over{x+\xi-i0}}\Biggr).
\end{equation}

Similar results were obtained in a recent paper~\cite{Diehl}.

\end{document}